\documentclass[cameraready]{Interspeech}

\title{ProSDD: Learning Prosodic Representations for Speech Deepfake Detection against Expressive and Emotional Attacks}
\author[affiliation={1},orcid=0009-0001-3057-2021]{Aurosweta}{Mahapatra}
\author[affiliation={1}, orcid=0000-0003-4593-9057]{Ismail Rasim}{Ulgen}
\author[affiliation={2}, orcid=0000-0001-9133-3000]{Kong Aik}{Lee}
\author[affiliation={3}, orcid=0000-0002-6097-9164]{Nicholas}{Andrews}
\author[affiliation={1}, orcid=0000-0001-8078-3305]~{Berrak~Sisman}{}

\address{
    $^1$ Center for Language and Speech Processing (CLSP), Johns Hopkins University, USA \\
     $^2$ Hong Kong Polytechnic University, Hong Kong \\
      $^3$ Human Language Technology Center of Excellence (HLT COE), Johns Hopkins University, USA}

\email{amahapa2@jhu.edu, noa@jhu.edu, sisman@jhu.edu}

\keywords{Speech deepfake detection, prosody, emotion}

\usepackage{comment}
\usepackage{amsmath}
\usepackage{booktabs}
\usepackage{multirow}
\usepackage{cite}
\usepackage{booktabs}
\usepackage{subcaption}
\usepackage{makecell}
\usepackage{graphicx}
\usepackage{enumitem}
\usepackage{csquotes}
\usepackage{url}
\usepackage{hyperref}

\begin{document}

\maketitle
\begin{abstract}

Speech deepfake detection (SDD) systems perform well on standard benchmarks datasets but often fail to generalize to expressive and emotional spoofing attacks. Many methods rely on spoof-heavy training data, learning dataset-specific artifacts rather than transferable cues of natural speech. In contrast, humans internalize variability in real speech and detect fakes as deviations from it. We introduce ProSDD, a two-stage framework that enriches model embeddings through supervised masked prediction of speaker-conditioned prosodic variation based on pitch, voice activity, and energy. Stage I learns prosodic variability from real speech, and Stage II jointly optimizes this objective with spoof classification. ProSDD consistently outperforms baselines under both ASVspoof 2019 and 2024 training, reducing ASVspoof 2024 EER from 25.43\% to 16.14\% (2019-trained) and from 39.62\% to 7.38\% (2024-trained), while achieving ~50\% relative reductions on EmoFake and EmoSpoof-TTS.

\end{abstract}


\begin{figure*}[t]
\centering
\begin{subfigure}[t]{0.47\textwidth}
    \centering
    \includegraphics[width=0.85\linewidth]{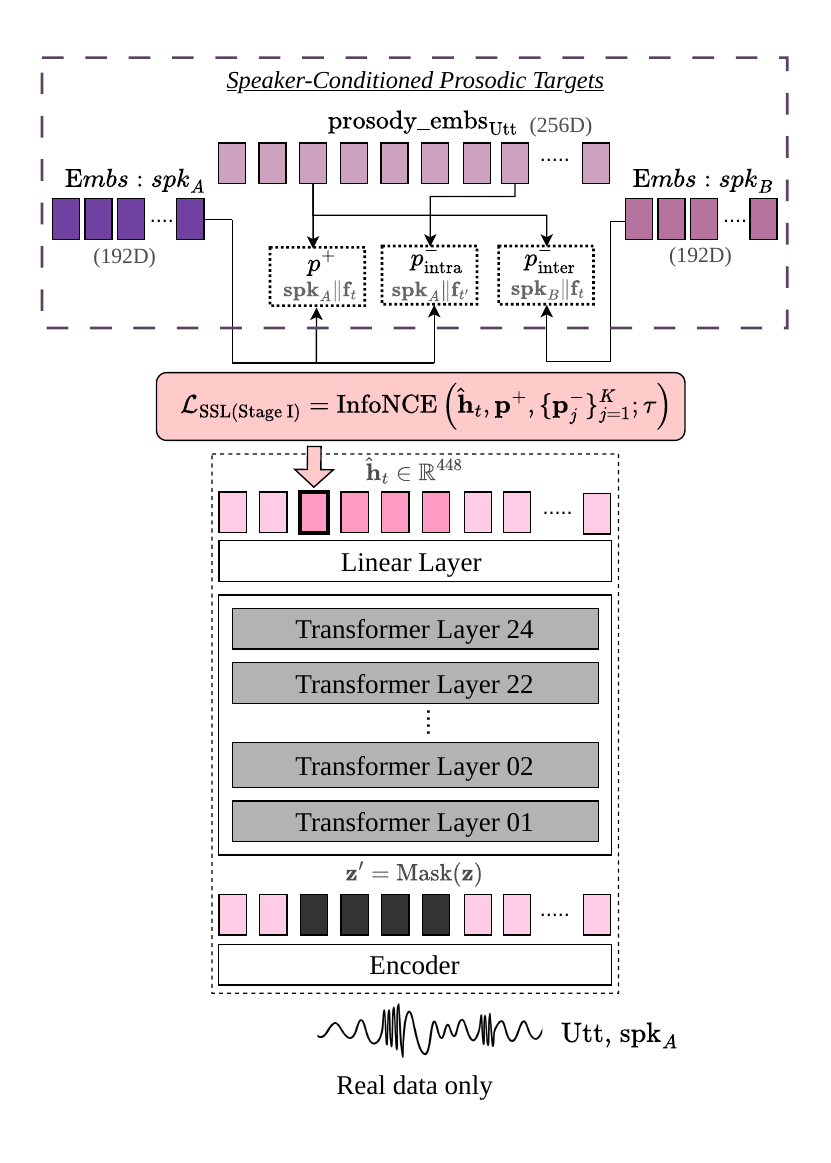}
    \vspace{-3mm}
    \caption{Stage I}
    \label{fig:stage1}
\end{subfigure}\hspace{-0.04\textwidth}
\begin{subfigure}[t]{0.56\textwidth}
    \centering
    \includegraphics[width=0.95\linewidth]{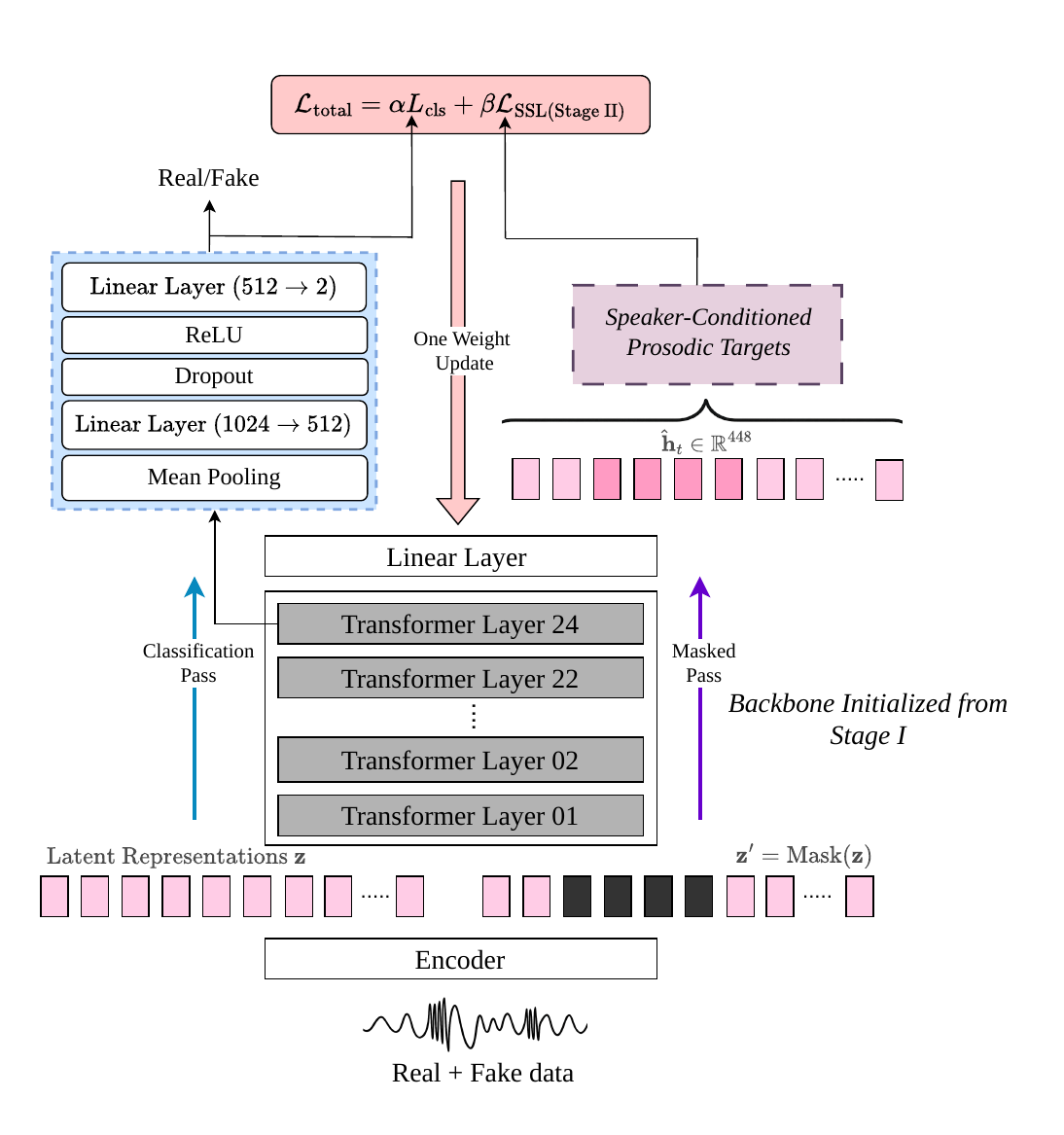}
    \vspace{-3mm}
    \caption{Stage II}
    \label{fig:stage2}
\end{subfigure}
\vspace{-3mm}
\caption{Two-stage training framework of ProSDD. Stage I learns speaker-conditioned prosodic representations from real speech, and Stage II jointly optimizes spoof classification with supervised masked prediction.}
\label{fig:prosd_arch}
\vspace{-4mm}
\end{figure*}
\section{Introduction}

Speech deepfake detection (SDD) aims to distinguish synthetic speech generated by text-to-speech (TTS) and voice conversion (VC) systems from genuine human speech \cite{AS_Survey,SDD_survey}. As synthesis models advance in naturalness, speaker similarity, and emotional expressiveness, this task becomes increasingly challenging \cite{emotionAS, EmoFake}. Over the past decade, SDD has advanced across diverse attack types, channel conditions, adversarial settings, multilingual scenarios \cite{mlaad}, and in-the-wild environments \cite{inthewild}. This progress has been driven by community benchmarks such as ASVspoof \cite{ASVspoof2015} and the ADD challenges \cite{ADD2022, ADD2023}, which introduced widely used datasets including ASVspoof versions \cite{ASVspoof2019, ASVspoof2021, ASVspoof2024}. While current SDD systems show strong performance on benchmark datasets, their robustness often degrades on emotionally rich speech beyond the training distribution \cite{EmoFake, emotionAS, mahapatra2025hula}. 

Current SDD approaches include end to end waveform models \cite{RawNet2, AASIST}, handcrafted spectral features with supervised classifiers \cite{LCNN, ASSERT, ResNet34}, and self supervised learning backbones with a classifier \cite{ssl_sls, ssl_rawnet2}. SSL based systems dominate due to large scale pretraining and flexible fine tuning, yet their robustness to emotional and expressive speech remains limited \cite{mahapatra2025hula}. Fine tuning is typically driven only by a spoof classification objective, often coupled with increasingly complex classifiers. This encourages reliance on spoof specific artifacts rather than structured properties of natural speech, such as prosodic variation, limiting generalization under distribution shifts. Recent studies highlight the vulnerability of existing SDD models to emotionally expressive synthetic attacks \cite{emotionAS, mahapatra2025hula, EmoFake}. Other work shows that incorporating emotion-related and prosodic cues can improve detection performance \cite{pitchimperfect, AS_SER, ASV_prososdy_AS, li2025emoanti}. However, limited work integrates expressive and emotional cues within SSL-based deepfake detectors to enhance generalization while maintaining strong performance on standard benchmarks.



This paper identifies two key limitations in current training paradigms. First, although modern SDD systems leverage pretrained representations, downstream fine-tuning for spoof detection typically relies on a classification objective over datasets primarily composed of spoofed samples. This may bias learning toward spoof- and dataset-specific artifacts rather than understanding transferable cues in natural speech limiting generalization. Second, while modern synthesis systems produce highly natural and expressive speech, they often exhibit subtle prosodic inconsistencies \cite{promismatch}. From a human perceptual perspective, listeners are not exposed to diverse synthetic attacks; instead, they internalize variability in authentic speech, particularly in prosodic and speaker-level patterns, and detect synthetic as deviations from this structure \cite{humanperception-DF,human_perception_subjetive_eval}. We hypothesize that incorporating such perceptual cues into model can improve SDD generalization. Consistent with this intuition, modeling prosodic structure and leveraging real speech have shown promise in SDD \cite{slim, ASV_prososdy_AS,mahapatra2025hula}, although such approaches remain relatively underexplored.

This paper proposes a two-stage supervised masked prediction framework that enriches model embeddings through speaker-conditioned prosodic variation modeled using pitch ($F_0$), voice activity, and energy. Stage I learns structured prosodic representations from real speech, and Stage II retains the masked prediction objective as auxiliary supervision alongside spoof classification. This design improves robustness to expressive and emotional attacks while maintaining competitive performance on standard benchmarks. Our contributions are: 
\begin{itemize}
    \item We introduce ProSDD, a two-stage supervised masked prediction framework that structures model representations through speaker-conditioned prosodic variation to enhance generalization in spoof detection
    \item We show that learning structured prosodic variation from real speech prior to spoof classification improves generalization to expressive and emotional synthetic speech.
    \item We demonstrate that enriched backbone representations enable strong cross-domain performance without relying on complex classifier architectures.
\end{itemize}
We publicly release ProSDD to support reproducibility. \footnote{\url{https://prosdd.github.io/ProSDD_website/}}

\section{Related Work}
Self-supervised learning (SSL) backbones are widely used for speech deepfake detection due to strong performance \cite{ssl_sls, ssl_rawnet2, ssl-mamba}. However, their robustness to emotional and expressive synthetic speech remains limited. Studies show that state-of-the-art SDD systems exhibit performance variations depending on emotion that can be exploited through targeted attacks \cite{EmoFake, emotionAS}, and fine-tuning on emotional speech often fails to generalize across domains and benchmarks \cite{mahapatra2025hula}. Some approaches incorporate emotion-related representations, such as speech emotion recognition (SER) features, to improve robustness \cite{AS_SER, li2025emoanti}. Yet these methods typically treat prosodic or emotion cues as additional classifier inputs, rather than explicitly enriching SSL representations with expressive variability.

Prosody plays a central role in expressive speech synthesis and perception. Modern TTS \cite{Emoqtts, expressive_glowtts, F5TTS, Emo-ctrlTTS} and voice conversion systems \cite{expressive_vc2, expressive_vc3, expressive_vc5} rely on pitch, energy, and voice activity to convey emotion and speaking style. Prosodic variation across speakers and utterances \cite{prosody, prososdy2, prospkvariation} makes accurate modeling challenging. There is evidence that human listeners are sensitive to subtle prosodic inconsistencies in synthetic speech and leverage them for deepfake detection \cite{humanperception-DF, human_perception_subjetive_eval}. Despite this, prosody remains underutilized in SDD. Prior work explores pitch cues \cite{pitchimperfect, produr_sdd}, voiced–unvoiced patterns \cite{VUV}, fusion with speaker embeddings \cite{ASV_prososdy_AS}, or prosody-aware SSL adaptations \cite{mahapatra2025hula}, typically incorporating prosodic features within the classification pipeline or underutilizing broader prosodic variation. In contrast, we use speaker-conditioned prosodic variation as a supervisory signal that directly structures the SSL backbone via supervised masked prediction, encouraging internalization of natural prosodic variability before spoof discrimination.

\section{Proposed Method}
We introduce \textbf{ProSDD}, a two-stage speech deepfake detection framework that enriches the contextual representations of a pretrained SSL backbone through supervised modeling of speaker-conditioned prosodic variation. Figure~\ref{fig:prosd_arch} illustrates the overall architecture. In Stage I, the backbone is fine-tuned using only real speech with a supervised masked prediction objective that learns structured prosodic representations conditioned on speaker identity. This stage encourages the model to internalize natural prosodic variability before exposure to spoofed speech. In Stage II, the Stage I weights initialize spoof detection training. Spoof classification is jointly optimized with the same masked prosodic objective used in Stage I, which now serves as an auxiliary task. Training is performed using two forward passes per step: a masked pass for prosodic supervision and an unmasked pass for spoof classification. During inference, only the XLS-R backbone and a lightweight classifier head are used. 

\subsection{Stage I: Prosody-Driven Representation Learning (Real Speech Only)}

Stage I fine-tunes the pretrained XLS-R on real speech to explicitly encode speaker-conditioned prosodic structure.

\subsubsection{Construction of Speaker-Conditioned Prosodic Targets}

The masked prediction targets consist of a speaker embedding and a frame-level prosodic embedding.

\textbf{Speaker embedding.} We extract 192-dimensional utterance-level embeddings using a pretrained ECAPA-TDNN model \cite{ecapa}. For each speaker, embeddings from all available utterances are averaged and L2-normalized to obtain a single speaker representation $\mathbf{spk} \in \mathbb{R}^{D_s}$, where $D_s = 192$.

\textbf{Prosodic embedding.} Frame-level prosodic embeddings $\mathbf{f}_t \in \mathbb{R}^{D_p}$, where $D_p = 256$ are extracted using the prosody encoder following \cite{MPM-Pro}, which integrates pitch ($F_0$), voice activity, and energy to capture fine-grained temporal variation.

For an utterance with $T$ frames, the speaker-conditioned prosodic target sequence is constructed as

\begin{equation}
\mathbf{P}^{\text{target}} =
\left[
\mathbf{spk} \Vert \mathbf{f}_1,\,
\mathbf{spk} \Vert \mathbf{f}_2,\,
\dots,\,
\mathbf{spk} \Vert \mathbf{f}_T
\right]
\in \mathbb{R}^{T \times D},
\end{equation}

\noindent where $\Vert$ denotes concatenation and $D=D_s + D_p = 448$. Repeating the speaker embedding across frames allows each masked position to be predicted with respect to both speaker identity and local prosodic variation.

\subsubsection{Supervised Masked Prediction Objective}
\textit{Why Supervised Masked Prediction?} SSL backbones such as XLS-R are pretrained with a self-supervised masked prediction objective, learning contextual speech representations. These models perform well for spoof-detection when fine-tuned with a classification objective. However, their generalization to emotional and expressive speech remains limited. Synthetic speech often contains subtle prosodic inconsistencies, which humans exploit for detection. We therefore enrich pretrained embeddings with explicit speaker-conditioned prosodic structure through a supervised masked prediction objective. Conditioning on speaker embeddings encourages the model to capture both within utterance prosodic patterns and variability across speakers, improving robustness under expressive conditions. To implement this objective, we proceed as follows. 

Let $z$ denote the latent representations produced by the convolutional encoder and projection layers of XLS-R. Span masking is applied to obtain the masked sequence $z' = \mathrm{Mask}(z)$. The masked sequence is passed through the Transformer stack to obtain contextualized embeddings $\mathbf{h}_t \in \mathbb{R}^{1024}$.

A linear projection layer maps these embeddings to the 448-dimensional target space:
\[
\hat{\mathbf{h}}_t = \mathrm{Linear}_{1024 \rightarrow 448}(\mathbf{h}_t).
\]

\noindent For a masked frame $t$ belonging to speaker $\mathbf{spk}_A$, the positive target is

\[
\mathbf{p}^{+} = \mathbf{spk}_A \Vert \mathbf{f}_t.
\]

\noindent Two types of negatives are sampled:

\begin{enumerate}[label=(\roman*)]
    \item \textbf{Intra-speaker negatives} \textit{(same speaker, different prosody)}:
    \[
    \mathbf{p}^{-}_{\text{intra}} = \mathbf{spk}_A \Vert \mathbf{f}_{t'}, 
    \quad t' \neq t \textit{ ; same utterance}
    \]

    \item \textbf{Inter-speaker negatives:} \textit{(different speaker, same prosody)}
    \[
    \mathbf{p}^{-}_{\text{inter}} = \mathbf{spk}_B \Vert \mathbf{f}_{t}
    \]
\end{enumerate}

\noindent where $\mathbf{spk}_B$ corresponds to a different speaker in the batch.

For each masked frame, $K$ negatives are sampled (half intra-speaker and half inter-speaker), where $K=100$ . The objective is formulated using an InfoNCE loss:

\begin{equation}
\mathcal{L}_{\text{SSL(Stage I)}} =
\mathrm{InfoNCE}
\left(
\hat{\mathbf{h}}_t,
\mathbf{p}^{+},
\{\mathbf{p}^{-}_j\}_{j=1}^{K};
\tau
\right),
\end{equation}

\noindent where cosine similarity is used and $\tau$ is a temperature hyperparameter. The contrastive loss differentiates correct and incorrect speaker–prosody pairs, promoting structured modeling of prosodic variation and explicitly enriching the contextual embeddings with prosodic structure.

\subsection{Stage II: Spoof Classification with Prosodic Auxiliary Supervision}

Stage II initializes the backbone with Stage I weights and trains on spoof detection data. The masked prediction objective is retained as an auxiliary task to preserve structured prosodic modeling while learning to discriminate real and synthetic speech.

\subsubsection{Two-Pass Training Strategy}

Each training step consists of two forward passes followed by a single parameter update.

\begin{enumerate}[label=(\roman*)]
\item \textbf{Masked pass.}
Span masking is applied to the latent representations, and the supervised masked prediction loss defined in Stage I is computed.

\item \textbf{Classification pass.}
The unmasked latent representations are passed through the Transformer to obtain contextual embeddings $\mathbf{h}_t \in \mathbb{R}^{1024}$. These embeddings are mean-pooled over time and fed to a classifier to compute the spoof classification loss.
\end{enumerate}

\noindent Using separate masked and unmasked passes prevents the classifier from relying on partially reconstructed representations during early training.

\subsubsection{Overall Training Objective}

The total loss in Stage II is

\begin{equation}
\mathcal{L}_{\text{total}} =
\alpha L_{\text{cls}}
+
\beta \mathcal{L}_{\text{SSL(Stage II)}},
\end{equation}

\noindent where $L_{\text{cls}}$ is the weighted cross-entropy loss for spoof classification, and $\mathcal{L}_{\text{SSL(Stage II)}}$ is the masked prediction loss for Stage~II, following Eq.~(2). The coefficients $\alpha$ and $\beta$ balance spoof discrimination and prosodic supervision.

\subsubsection{Lightweight Classifier Head}

To emphasize the contribution of enriched SSL representations rather than classifier complexity, we employ a lightweight classifier consisting of a linear layer, dropout, ReLU activation, and a final linear layer. We intentionally avoid attention pooling and complex architectural designs to ensure that performance gains are attributable to prosody-driven representation learning.


During inference, only the XLS-R backbone and the classifier head obtained after Stage II are used.

\begin{table*}[t]
\centering
\renewcommand{\arraystretch}{1.1}
\setlength{\tabcolsep}{4pt}

\caption{Comparison on Standard and Emotional Benchmarks under Two Training Settings (\% EER $\downarrow$)}
\label{tab:bothtrain}
\vspace{-3mm}
\begin{subtable}[t]{0.485\textwidth}
\centering
\caption{Trained on ASVspoof 2019 LA.}
\label{tab:train2019}
\begin{tabular}{p{2cm}|ccc|c|c}
\hline
\multirow{2}{*}{Models} & \multicolumn{3}{c|}{ASVspoof} & \multirow{2}{*}{EmoFake} & \multirow{2}{*}{EmoSpoof} \\
\cline{2-4}
 & 2019 & 2021 & 2024 &  &  \\
\hline
RawNet2    & 4.60 & 8.08 & 40.67 & 21.71 & 43.04 \\
\hline
AASIST    & 0.83 & 8.15 & 35.53 & 13.64 & 31.06 \\
\hline
XLSR-SLS   & 0.56 & \textbf{3.04} & 25.43 & 8.84 & 18.92 \\
\hline
\textbf{ProSDD} & \textbf{0.42} & 3.87 & \textbf{16.14} & \textbf{3.70} & \textbf{9.54} \\
\hline
\end{tabular}
\end{subtable}
\hfill
\begin{subtable}[t]{0.485\textwidth}
\centering
\caption{Trained on ASVspoof 2024.}
\label{tab:train2024}
\begin{tabular}{p{2cm}|ccc|c|c}
\hline
\multirow{2}{*}{Models} & \multicolumn{3}{c|}{ASVspoof} & \multirow{2}{*}{EmoFake} & \multirow{2}{*}{EmoSpoof} \\
\cline{2-4}
& 2019 & 2021 & 2024 &  &  \\
\hline
RawNet2    & 24.75 & 25.59 & 43.61 & 49.49 & 27.13 \\
\hline
AASIST     & 23.16    & 22.74    & 25.77   & 62.71    & 15.19    \\
\hline
XLSR-SLS   & 27.00 & 26.54 & 39.62 & 58.57 & 25.92 \\
\hline
\textbf{ProSDD} & \textbf{19.04} & \textbf{18.08} & \textbf{7.38} & \textbf{25.06} & \textbf{11.96} \\
\hline
\end{tabular}
\end{subtable}

\renewcommand{\arraystretch}{1.0}
\vspace{-3mm}
\end{table*}

\section{Experimental Setup}
We describe the datasets, baselines, and implementation details used to evaluate ProSDD.

\textbf{Dataset.} We group data into training, standard benchmarks, and emotional/expressive benchmarks. Stage~I training uses LibriSpeech train-clean-100 and dev (bona fide only) \cite{librispeech}. Stage~II training uses ASVspoof~2019 LA train/dev \cite{ASVspoof2019} and ASVspoof~2024 train/dev \cite{ASVspoof2024}. For traditional benchmarks, we evaluate on ASVspoof~2019 LA and ASVspoof~2021 LA \cite{ASVspoof2019,ASVspoof2021}. For emotional and expressive evaluation, we use EmoFake \cite{EmoFake}, EmoSpoof-TTS (abbreviated as EmoSpoof in Tables) \cite{emotionAS}, and ASVspoof~2024 Track~1 \cite{ASVspoof2024}, which includes modern expressive synthesis systems ~\cite{expressive_VITS, expressive_xtts, expressive_fastpitch, expressive_glowtts, expressive_gradtts, expressive_starganv2, expressive_diffVC, expressive_VAE_GAN}. All experiments follow the official train/dev/eval splits of each dataset. 

\textbf{Baseline.} We compare ProSDD with widely used and publicly available end-to-end and SSL-based baselines known for strong performance on traditional benchmarks: RawNet2 \cite{RawNet2}, AASIST \cite{AASIST}, and XLSR-SLS \cite{ssl_sls}. For ASVspoof~2019 LA training, we use official pretrained checkpoints for RawNet2/AASIST and train XLSR-SLS following \cite{ssl_sls}. For ASVspoof~2024 Track~1 training, all baselines are trained from scratch using their published protocols. 

\textbf{ProSDD Implementation Details.} Training is performed in two stages for 50 epochs with batch size 64 using fixed 4-second segments. Prosodic targets are fixed to 200 frames to match the SSL prediction tokens. In Stage~I, span masking uses length 8, masking probability 0.25, and temperature $\tau{=}0.07$. In Stage~II, masking probability is reduced to 0.15 with $\tau{=}0.1$. The joint loss uses $\alpha{=}1$; $\beta$ is set to 0.2 for the first 4 epochs and then reduced to 0.05 so that prosodic supervision acts as a regularizer while prioritizing spoof classification. RawBoost (Method 3) \cite{rawboost} is applied in Stage II, and XLSR-SLS \cite{ssl_sls} follows its original protocol using the same augmentation. Layer-wise learning rates are $1{\times}10^{-6}$ (SSL backbone), $1{\times}10^{-4}$ (projection), and $1{\times}10^{-5}$ (classifier), with weight decay $1{\times}10^{-4}$. For ASVspoof 2019 LA, model selection is based on training loss, as the development set does not introduce additional attack variability. For ASVspoof 2024, validation accuracy is used due to unseen attack types in the development set. 


\section{Results}
We evaluate ProSDD on traditional benchmarks and challenging emotional/expressive datasets to assess both in-domain performance and cross-domain generalization. 


\textbf{Performance on Traditional Benchmarks.} 
Table~\ref{tab:bothtrain} shows that ProSDD maintains strong performance on ASVspoof 2019 and 2021 under both training settings. When trained on ASVspoof 2019, ProSDD achieves 0.42\% equal error rate (EER) on ASVspoof 2019, outperforming XLSR-SLS (0.56\%), while remaining competitive on ASVspoof 2021 (3.87\% vs.\ 3.04\%). When trained on ASVspoof 2024 (TTS-only), ProSDD still generalizes effectively to ASVspoof 2019 and 2021, which include both TTS and VC attacks as well as channel variability. This indicates robustness beyond the training distribution.

\textbf{Emotion and Expressiveness.} 
As shown in Table~\ref{tab:bothtrain}, ProSDD substantially improves robustness on expressive datasets. When trained on ASVspoof 2019, it reduces EER on EmoFake from 8.84\% (XLSR-SLS) to 3.70\%, and on EmoSpoof-TTS from 18.92\% to 9.54\%, while achieving 16.14\% on ASVspoof 2024. When trained on ASVspoof 2024, ProSDD attains 7.38\% on the ASVspoof 2024 test set compared to 39.62\% for XLSR-SLS, and further achieves 11.96\% on EmoSpoof-TTS and 25.06\% on EmoFake. This setting is particularly challenging since EmoFake contains only VC attacks while training uses only TTS samples. Despite this cross-attack mismatch, ProSDD maintains strong performance, demonstrating robust generalization under expressive and attack-shift conditions. The gains are consistent across both training regimes, indicating that improvements are not dataset-specific. This supports our claim that explicitly modeling speaker conditioned prosodic structure strengthens representations in a way that generalizes across expressive and cross attack distribution shifts.

\begin{table}[t]
\centering
\setlength{\tabcolsep}{3pt}
\renewcommand{\arraystretch}{1.05}
\caption{Ablation study (trained on ASVspoof 2019 LA). 
MP denotes the supervised masked prediction objective. 
\enquote{w/o MP-SI} removes both (i) real-only prosodic pretraining in Stage I and (ii) masked prediction in Stage II. \enquote{w/o Stage I} removes real-only prosodic pretraining while retaining MP in Stage II (\% EER $\downarrow$).}

\label{tab:ablation}
\begin{tabular}{lccccc}
\toprule
Model & ASV19 & ASV21 & ASV24 & EmoFake & EmoSpoof \\
\midrule
w/o MP-SI      & 6.78 & 25.18 & 28.12 & 14.02 & 10.02 \\
w/o Stage I & 5.14 & 7.83  & \textbf{15.55} & 6.37 & 15.02 \\
\midrule
\textbf{ProSDD} & \textbf{0.42} & \textbf{3.87} & 16.14 & \textbf{3.70} & \textbf{9.54} \\
\bottomrule
\end{tabular}
\renewcommand{\arraystretch}{1.0}
\vspace{-6mm}
\end{table}
\vspace{0.3mm}


\textbf{Impact of Supervised Masked Prediction and Real-Only Pretraining.} Table~\ref{tab:ablation} isolates the roles of masked prediction (MP) and Stage~I pretraining. Removing MP and Stage I pretraining (w/o MP-SI) severely degrades performance on all benchmarks (e.g., 6.78\% vs.\ 0.42\% on ASVspoof 2019), indicating poor generalization to both standard and expressive conditions. Although this variant performs similarly on EmoSpoof-TTS, it drops notably on EmoFake and traditional benchmarks, revealing sensitivity to distribution shifts. Retaining MP only in Stage~II (w/o Stage~I) provides partial gains, particularly on ASVspoof 2024, but lacks consistent cross-dataset robustness. In contrast, the full two-stage ProSDD achieves the most stable performance across all settings. This highlights the importance of the strategy used to incorporate prosodic information. Related work \cite{ASV_prososdy_AS} provides prosodic and speaker cues to the classifier at the feature level, whereas ProSDD integrates speaker-conditioned prosodic structure directly into the SSL backbone representations via supervised masked prediction. Stage~I learns natural prosodic variability before spoof discrimination, and Stage~II preserves it through auxiliary supervision. Together, these results show that real-only prosodic pretraining with joint supervision is critical to enhance generalization.


\section{Conclusion}
We introduced ProSDD, a two-stage speech deepfake detection framework that first learns structured speaker-conditioned prosodic representations from bona fide speech via supervised masked prediction, and then jointly optimizes spoof classification with prosodic supervision as an auxiliary task. ProSDD substantially improves robustness for emotional and expressive datasets while maintaining competitive performance on traditional benchmarks under both ASVspoof 2019 and ASVspoof 2024 training settings. Ablation results confirm that real-only prosodic pretraining is critical for generalization, as Stage I embeds structured variability that guides spoof discrimination under distribution shifts. Overall, our findings demonstrate that explicitly modeling natural prosodic variability is key to building SDD systems that generalize beyond standard benchmarks.

\section{Acknowledgments}
This work was supported by:
\begin{itemize}
    \item The National Science Foundation (NSF) CAREER Award IIS-2533652.
    \item The Office of the Director of National Intelligence (ODNI), Intelligence Advanced Research Projects Activity (IARPA), via the ARTS Program, Contract \#D2023-2308110001.
\end{itemize}
The views and conclusions contained herein are those of the authors and should not be interpreted as necessarily representing the official policies, either expressed or implied, of ODNI, IARPA, or the U.S. Government. The U.S. Government is authorized to reproduce and distribute reprints for governmental purposes notwithstanding any copyright annotation therein.

\section{Generative AI Use Disclosure}
Generative AI tools were employed solely for language polishing of text written by the authors. These tools were not used to generate scientific content, results, experimental designs, analyses, or conclusions. All authors are responsible for the full content of this paper and consent to its submission.



\bibliographystyle{IEEEtran}
\bibliography{mybib}

\end{document}